\documentclass[12pt]{article}

\author{Yu.~M.~Zinoviev
       \thanks{E-mail address: yurii.zinoviev@ihep.ru} \\
        {\it Institute for High Energy Physics} \\
        {\it Protvino, Moscow Region, 142280, Russia}}
\title{On massive spin 2 interactions}

\date{}

\begin{document}

\maketitle

\begin{abstract}
In this paper we use a constructive approach based on gauge invariant
description of massive high spin particles for investigation of
possible interactions of massive spin 2 particle. We work with
general case of massive spin 2 particle living in constant curvature
$(A)dS_d$ background, which allows us carefully consider all flat
space, massless or partially massless limits. In the linear
approximation (cubic terms with no more than two derivatives in the
Lagrangians and linear terms with no more than one derivative in
gauge transformations) we investigate possible self-interaction,
interaction with matter (i.e. spin 0, 1 and 1/2 particles) and
interaction with gravity. 
\end{abstract}

\thispagestyle{empty}
\newpage
\setcounter{page}{1}

\section*{Introduction}

In all investigations of massless particles interactions gauge
invariance plays a crucial role. Not only it determines a kinematic
structure of free theory and guarantees a right number of physical
degrees of freedom, but also to a large extent fixes possible
interactions of such particles. This leads, in particular, to
formulation of so called constructive approach for investigation
of massless particles models \cite{OP65,FF79,MUF80,Wald86,BFPT06}. In
such approach, starting with appropriate collection of free massless
particles and requiring conservation of (modified) gauge invariance
in the process of switching on an interaction, one can consistently
reproduce such physically important theories as Yang-Mills, gravity
and supergravity.

Usual description of massive particles does not include gauge
invariance and it is hard to formulate one simple principle for
constructing consistent theories with such particles. A lot of
different requirements, such as conservation of right number of
physical degrees of freedom, smooth massless limit, tree level
unitarity and causality, was used in the past 
\cite{HS82,FPT92,CPD94,DPW00,DW01d,BKP99,BGKP99,BGP00}.

There exist two classes of consistent models for massive high spin
particles, namely, for massive non-abelian spin 1 particles and for
massive spin 3/2 ones. In both cases masses of gauge fields appear
as a result of spontaneous gauge symmetry breaking. One of the main
ingredients of this mechanism is the appearance of Goldstone
particles with non-homogeneous gauge transformations. This, in turn,
leads to the gauge invariant description of such massive spin 1 and
spin 3/2 particles. 
But such gauge invariant description of massive particles could be
constructed for higher spins as well, e.g.
\cite{Zin83,KZ97,Zin01,AGS02,Ham05,BHR05,HW05,BK05,BKL06,Met06}
This allows one to extend the above mentioned constructive approach
to the case of massive high spin particles. 

In the first section of our paper we give two simple examples of
such constructive approach. One of them deals with triplet of massive
spin 1 particles with gauge group $SU(2) \simeq O(3)$. We show that
constructive approach allows one to reproduce two well known
possibilities: one based on the non-linear $\sigma$-model \cite{FS71},
and, with the help of introduction of additional scalar field, usual
model of spontaneous symmetry breaking with the doublet of Higgs
fields. Another example devoted to electromagnetic interaction for
massive spin 3/2 particle \cite{DPW00,DW01d}. In the first linear
approximation (i.e. cubic terms in the Lagrangian and linear terms in
the gauge transformations) we construct the most general gauge
invariant Lagrangian. Our results unambiguously show that any model
with minimal gauge interaction of massive spin 3/2 particle in flat
space must be a part of some (spontaneously broken) supergravity
theory.

Main part of our paper devoted to massive spin 2 particle
interactions. It is an old but still unsolved problem and the main
difficulty is connected with the so called sixth degree of freedom
\cite{BD72}. Indeed, for the free particle it is easy to choose a
structure of mass terms so that there are exactly five degrees of
freedom in the model. But in the interacting theory this dangerous
(because it is a ghost) sixth degree of freedom reappear. Note that
it is, in principle, possible that in the full non-linear theory this
DoF become physical \cite{BG02,Pet05}, but in what follows we will
not consider such possibility. One more difficulty that appears in
such theories is the absence of smooth massless limit
\cite{Zak70,DV72}. Moreover, when one considers such theory in a
background space with non-zero cosmological term (de Sitter or anti
de Sitter spaces) one faces an ambiguity between flat space and
massless limits \cite{KMP00,Por00,DDGV02}.

Here we use constructive approach based on the gauge invariant
description of massive particles for the investigation of possible
massive spin 2 interactions. It turns out that some results depends
on the dimensionality of space-time as well as on the presence (or
absence) of cosmological term. So we start in section 2 with the
gauge invariant formulation of free massive spin 2 particle in the
space-time of arbitrary $d > 2$ dimension with non-zero cosmological
term, that could be positive or negative. This includes, in
particular, the so called partially massless spin 2 particle, which
could exist in the de Sitter space only
\cite{DW01,DW01a,DW01c,Zin01,SV06}. 

Then, in section 3 we consider possible self-interaction of such
particles in linear approximation. We will not make any suggestion
about gauge algebra which could stand behind such theory nor we will
not insist on any possible geometrical interpretation. Instead, we
use a "brute force" method for our construction. Namely, we write the
most general Lorentz invariant cubic vertexes in the Lagrangian as
well as the most general ansatz for gauge transformations linear in
fields and require that the Lagrangian will be gauge invariant. The
only restriction (besides Lorentz invariance) we impose is
the restriction on the number of derivatives. In this paper we
consider only interactions with no more than two derivatives (and
gauge transformations with no more than one derivative) leaving
investigation of possible higher derivatives interactions for the
future. 

Clearly, the existence of linear approximation does not guarantee
that the construction of full non-linear theory is possible because
obstruction can appear in the next approximations. Nevertheless, the
investigation of linear approximation is a very important step,
because structure of this approximation (and its whole existence) does
not depend on the presence of any other field in the theory. Only in
the next quadratic approximation one faces the problem that the
closure of gauge algebra requires the presence of right (finite or
infinite) collection of fields. Thus, the results obtained in our
paper are essentially model independent (up to restriction on the
number of derivatives).

For the generic values of mass and cosmological constant the linear
approximation for self-interaction exists in any space-time dimension
$d \ge 3$, but for partially massless spin 2 particle it exists in
$d=4$ dimensions only. This clearly related with the fact that only
in $d=4$ partially massless theories are conformally invariant
\cite{DW04}. One of the non-trivial results is that the structure of
gauge transformations for Goldstone field $A_\mu$ turns out to be
non-canonical, so if one tries to interpret (as usual in gravity)
the $\xi_\mu$ gauge transformations as general coordinate ones, then
this field should give some non-linear realization of this
transformations. One more non-trivial result is that two gauge
transformations --- vector $\xi_\mu$ and scalar $\lambda$ ones, do
not commute when the mass $m \ne 0$ forming an algebra similar to
Weyl one. 

In the next section we consider possible cubic couplings of massive
spin two particles with matter. Namely, we consider interaction with
(massive or massless) spin 0, spin 1 and spin 1/2 particles. In all
cases (except massless spin 1 in $d=4$) our results show an ambiguity
between flat and massless limits, which shows itself through
non-trivial dependence of coupling constant for scalar Goldstone
field $\varphi$ and matter fields interactions on mass and
cosmological constant. As a result, partially massless spin 2
particle can interact with matter having traceless energy-momentum
tensor only, the most important example being the coupling of
partially massless spin 2 with massless spin 1.

Our last section deals with possible interactions for massive spin 2
particles with gravity. As is well known, it is impossible to
construct non-trivial consistent interacting theory for collection of
massless spin 2 particles \cite{Wald86a,Wald86b,BDGH00,Bou02,Anc03},
but it still leaves the possibility to construct non-trivial
interaction of gravity with (one or many) massive spin 2 particles.
Indeed, such possibility attend much interest recently in very
different contexts. First of all, we show that it is impossible to
construct cubic vertex with two massless and one massive spin 2
particles while vertex with two massive and one massless ones does
exist. Moreover, it is easy to construct general covariant vertex
with two massive fields and arbitrary number of massless gravitons.
As for the interaction of partially massless spin 2 particles with
gravity, no restrictions on the space-time dimension arise in this
approximation.

\section{Toy models}

In this section we present two simple examples of constructive
approach to massive high spin particles interactions. First one deals
with the triplet of massive spin 1 particles, while second one
devoted to possible gauge interactions for massive spin 3/2 particle.
Results of this section where already known before, but it is
instructive to see how they can be reproduced using gauge invariant
formulation alone.

\subsection{Spin 1}

In this subsection we illustrate our approach by the simplest
non-abelian $SU(2) \sim O(3)$ gauge theory with triplet of vector
fields $A_\mu{}^a$, $a=1,2,3$. It is well known that with the help of
triplet scalar Goldstone fields $\varphi^a$ one can easily construct
gauge invariant description of free massive particles. In this, the
Lagrangian and gauge transformations look like:
\begin{eqnarray}
{\cal L}_0 &=& - \frac{1}{4} (A_{\mu\nu}{}^a)^2 + \frac{m^2}{2}
(A_\mu{}^a)^2 - m A_\mu{}^a \partial_\mu \varphi^a + \frac{1}{2}
(\partial_\mu \varphi^a)^2 \nonumber \\
\delta_0 A_\mu{}^a &=& \partial_\mu \xi^a,  \qquad
\delta_0 \varphi^a = m \xi^a
\end{eqnarray}
Now, if we switch on usual non-abelian gauge interaction:
\begin{eqnarray}
A_{\mu\nu}{}^a &=& \partial_\mu A_\nu{}^a - g \varepsilon^{abc}
A_\mu{}^b A_\nu{}^c - (\mu \leftrightarrow \nu) \nonumber \\
\delta_0 A_\mu{}^a &=& \partial_\mu \xi^a - g \varepsilon^{abc}
A_\mu{}^b \xi^c
\end{eqnarray}
then the Lagrangian will not be gauge invariant any more:
\begin{equation}
\delta_0 {\cal L}_0 = - g m \varepsilon^{abc} A_\mu{}^a \partial_\mu
\varphi^b \xi^c
\end{equation}
but at the linear approximation the gauge invariance could be easily
restored with additional vertex in the Lagrangian and appropriate
corrections to gauge transformations:
\begin{equation}
{\cal L}_1 = a_1 \varepsilon^{abc} A_\mu{}^a \varphi^b \partial_\mu
\varphi^c, \qquad \delta_1 \varphi^a = g_1 \varepsilon^{abc}
\varphi^b \xi^c
\end{equation}
provided $a_1 = - g/2$, $g_1 = - g/2!$. Note, that non-canonical
value $g/2$ in the gauge transformations of scalar fields $\varphi^a$
shows that these fields do not transform under linear triplet
representation of gauge group. As we already mentioned in the
introduction, this first step does not depend on the presence of any
other fields in the model. Now we can proceed in at least two
different ways. At first, we can evade introduction of any additional
fields and go on by introducing into the Lagrangian terms of
the type $A_\mu \varphi^2 \partial_\mu \varphi$, $\varphi^2
(\partial_\mu \varphi)^2$ and so on. It would lead us to the
essentially non-linear theory with the Lagrangian \cite{FS71}:
\begin{equation}
{\cal L} = - \frac{1}{4} (A_{\mu\nu}{}^a)^2 + \frac{m^2}{2}
(A_\mu{}^a)^2 - m A_\mu{}^a E^{ak}(\varphi) \partial_\mu \varphi^k +
\frac{1}{2} g^{kl}(\varphi) \partial_\mu \varphi^k \partial_\mu
\varphi^l
\end{equation}
and gauge transformations for scalar fields:
\begin{equation}
\delta \varphi^k = m (E^{-1})^{ka} \xi^a
\end{equation}
where $g^{kl} = E^{ak} E^{al}$. This Lagrangian will be gauge
invariant provided the "triad" $E^{ak}$ satisfy the equation
\begin{equation}
\frac{\partial E^{ak}}{\partial \varphi^l} - \frac{\partial E^{al}}
{\partial \varphi^k} = g \varepsilon^{abc} E^{bk} E^{cl}
\end{equation}
It is easy to check that non-canonical value $g/2$ in the linear
approximation for the $\varphi^a$ gauge transformations is a direct
consequence of this equation.

But there is another way. We can introduce one more scalar field
$\chi$ and try to stop iterations at some order. And indeed, if we
add to the Lagrangian of linear approximation all possible cubic and
quartic terms with new field:
\begin{equation}
{\cal L}_2 = \frac{1}{2} (\partial_\mu \chi)^2 + a_2 \chi (\partial
A)^a \varphi^a + a_3 \chi A_\mu{}^a \partial_\mu \varphi^a + a_4 m
\chi (A_\mu{}^a)^2 + a_5 (A_\mu{}^a)^2 \varphi^2 + a_6 \chi^2
(A_\mu{}^a)^2
\end{equation}
as well as appropriate corrections to the gauge transformations:
\begin{equation}
\delta \chi = g_2 (\varphi \xi), \qquad \delta_2 \varphi^a = g_3 \chi
\xi^a
\end{equation}
we can obtain full gauge invariant theory, provided:
$$
g_2 = a_2 = \frac{g}{2}, \qquad a_3 = g, \qquad g_3 = a_4 = -
\frac{g}{2}, \qquad a_5 = a_6 = \frac{g^2}{8}
$$
Moreover, if we introduce shifted variable 
$\tilde{\chi} = \chi - \frac{2m}{g}$, then total Lagrangian could be
rewritten in a familiar form:
\begin{eqnarray}
{\cal L} &=& - \frac{1}{4} (A_{\mu\nu}{}^a)^2 + \frac{1}{2}
(\partial_\mu \varphi^a)^2 + \frac{1}{2} (\partial_\mu \chi)^2 -
\frac{g}{2} \varepsilon^{abc} A_\mu{}^a \varphi^b \partial_\mu
\varphi^c + \nonumber \\
 && + \frac{g}{2} A_\mu{}^a (\tilde{\chi} \partial_\mu \varphi^a -
\partial_\mu \tilde{\chi} \varphi^a) + \frac{g^2}{8} (A_\mu{}^a)^2
(\varphi^2 + \tilde{\chi}^2)
\end{eqnarray}
Up to the arbitrary potential depending on the invariant combination
$\varphi^2 + \tilde{\chi}^2$ only this is just the well known model
for spontaneous gauge symmetry breaking through the Higgs mechanism.
In this, four scalar fields $\varphi^a$ and $\chi$ are transformed
under the linear doublet representation of gauge group.

Note, that one more possibility arises when one consider systems with
infinite number of gauge fields such as so called Kaluza-Klein models
e.g. \cite{DD84,CS85,Reu88,CZ92,TZ96b,Hoh05}. Indeed, let us consider
simplest example --- five-dimensional Yang-Mills theory with gauge
fields $A_\mu{}^a$, where $M = 0,1,2,3,5$, and some non-abelian gauge
algebra $[t^a, t^b] = f^{abc} t^c$. Suppose now that one dimension is
a compact one, being a circle of radius $R$. From a four-dimensional
point of view, such fields represent combinations of vector and
scalar ones: $A_M(x_\mu, x_5) \Rightarrow A_\mu(x, x_5), \varphi(x,
x_5)$, which equivalent to infinite number of four-dimensional fields:
\begin{eqnarray}
A_\mu{}^a(x_\mu, x_5) &=& \sum_{n=-\infty}^{n=\infty}
A_{\mu(n)}{}^a(x) \exp(iMnx_5) \nonumber \\
\varphi^a(x_\mu, x_5) &=& \sum_{n=-\infty}^{n=\infty}
\varphi_{(n)}{}^a(x) \exp(iMnx_5) 
\end{eqnarray}
where $M \simeq \frac{1}{R}$,  $A_{\mu(n)}^* = A_{\mu(-n)}$ and
similarly for $\varphi$. Performing integration on $x_5$, one obtains
four-dimensional gauge theory with infinite dimensional gauge algebra:
$$
 [ t_n{}^a, t_m{}^b ] = f^{abc} t_{n+m}{}^c
$$
In this, gauge transformations of all fields have the form:
\begin{eqnarray}
\delta A_{\mu(n)}{}^a &=& \partial_\mu \Lambda_n{}^a + f^{abc} 
\Lambda_k{}^b A_{\mu(n-k)}{}^c \nonumber \\
\delta \varphi_n{}^a &=& Mn \Lambda_n{}^a + f^{abc}
\Lambda_k{}^b \varphi_{(n-k)}{}^c
\end{eqnarray}
From the last line we see, that all the scalar fields, except
$\varphi_0$, have non-homogeneous transformation laws and play the
role of Goldstone fields. As a result, such model describes finite
number of massless vector $A_{\mu 0}{}^a$ and scalar 
$\varphi_0{}^a$ fields and infinite tower of massive $n \ne 0$ vector
fields. Introducing appropriate covariant derivatives for these
fields:
$$
D_\mu \varphi_n{}^a = \partial_\mu \varphi_n{}^a - f^{abc} 
A_{\mu(k)}{}^b \varphi_{(n-k)}{}^c - Mn A_{\mu(n)}{}^a
$$
the total Lagrangian could be written in a compact form:
$$
{\cal L} = \sum_{n=-\infty}^{n=\infty} [ - \frac{1}{4}
A_{\mu\nu(n)}{}^a A_{\mu\nu(-n)}{}^a + \frac{1}{2} D_\mu
\varphi_n{}^a D_\mu \varphi_{-n}{}^a ]
$$
More complicated examples, corresponding to partial breaking of
initial gauge algebra, could be found in \cite{TZ96b}.

\subsection{Spin 3/2}

Our second example devoted to the possible electromagnetic
interactions for massive spin 3/2 particles \cite{DPW00,DW01d}. As is
well known using two fields --- vector-spinor $\Psi_\mu$ and spinor
$\chi$ one can construct gauge invariant formulation of massive spin
3/2 particle. So we start with the sum of free massive spin 3/2 and
massless spin 1 Lagrangians:
\begin{eqnarray}
{\cal L}_0 &=& \frac{i}{2} \varepsilon^{\mu\nu\alpha\beta}
\bar{\Psi}_\mu \gamma_5 \gamma_\nu \partial_\alpha \Psi_\beta +
\frac{i}{2} \bar{\chi} \gamma^\mu \partial_\mu \chi - \frac{1}{4}
A_{\mu\nu}{}^2 - \nonumber \\
 && - \frac{m}{2} \bar{\Psi}_\mu \sigma^{\mu\nu} \Psi_\nu + i
\sqrt{\frac{3}{2}} m (\bar{\Psi} \gamma) \chi - m \bar{\chi} \chi
\end{eqnarray}
where gauge transformations leaving this Lagrangian invariant look
like:
\begin{equation}
\delta_0 \Psi_\mu = \partial_\mu \eta + \frac{im}{2} \gamma_\mu \eta,
\qquad \delta_0 \chi = \sqrt{\frac{3}{2}} m \eta, \qquad \delta A_\mu
= \partial_\mu \lambda
\end{equation}
Here $\eta$ is a spinor parameter, while $\lambda$ is a scalar. We
prefer to work with Majorana spinors, so in what follows all spinor
objects $\Psi_\mu$, $\chi$ and $\eta$ will be considered as doublets
of (anticommuting) Majorana spinors. In this, minimal gauge
interactions corresponds to the replacement of ordinary partial
derivatives by covariant ones:
\begin{equation}
\partial_\mu \Rightarrow D_\mu = \partial_\mu + q A_\mu, \qquad
q = \left( \begin{array}{cc} 0 & e \\ -e & 0 \end{array} \right),
\qquad q^2 = - e^2 I
\end{equation}
As a result of such replacement, the Lagrangian lost its gauge
invariance (just because covariant derivatives do not commute):
\begin{equation}
\delta_0 {\cal L}_0 = i \bar{\Psi}_\mu q \tilde{A}^{\mu\nu} \gamma_5
\gamma_\nu \eta, \qquad \tilde{A}^{\mu\nu} = \frac{1}{2}
\varepsilon^{\mu\nu\alpha\beta} A_{\alpha\beta}
\end{equation}
So we try to restore gauge invariance by adding to the Lagrangian
non-minimal terms linear in gauge field-strength $A_{\mu\nu}$:
\begin{eqnarray}
m {\cal L}_1 &=& \frac{1}{2} \bar{\psi}_\mu \left[ a_1 A^{\mu\nu} +
a_2 \gamma_5 \tilde{A}^{\mu\nu} + a_3 g^{\mu\nu} (\sigma A) + a_4
(A^{\mu\alpha} \sigma_\alpha{}^\nu + \sigma^{\mu\alpha}
A_\alpha{}^\nu) \right] q \Psi_\mu + \nonumber \\
 && + i \bar{\Psi}_\mu ( a_5 A^{\mu\nu} + a_6 \gamma_5
\tilde{A}^{\mu\nu}) \gamma_\nu q \chi + \frac{a_7}{2} \bar{\chi} q 
(\sigma A) \chi
\end{eqnarray}
as well as all possible linear terms for $\eta$-transformations for
all three fields:
\begin{eqnarray}
m \delta_1 \Psi_\mu &=& i q (\alpha_1 A_{\mu\nu} + \alpha_2 \gamma_5
\tilde{A}_{\mu\nu}) \gamma^\nu \eta \qquad m \delta_1 \chi = q
\alpha_3 (\sigma A) \eta \nonumber \\
m \delta_1 A_\mu &=& \alpha_4 (\bar{\Psi}_\mu q \eta) + i \alpha_5
(\bar{\chi} \gamma_\mu q \eta)
\end{eqnarray}
Straightforward calculations show that the gauge invariance could be
restored (up to terms quadratic in $A_{\mu\nu}$) provided all the
unknown coefficients are expressed in terms of two parameters, say
$\alpha_1$ and $\alpha_3$,
\begin{eqnarray*}
\alpha_2 &=& \alpha_1, \qquad \alpha_4 = 2 \alpha_1, \qquad \alpha_5 =
- 2 \alpha_3, \qquad a_1 = - 2 \alpha_1, \\
a_2 &=& 2 \alpha_1. \quad a_3 = a_4 = 0, \quad a_5 = a_6 = - 2
\alpha_3, \quad a_7 = 2 \sqrt{\frac{2}{3}} \alpha_3
\end{eqnarray*}
while this two parameters satisfy the relation:
\begin{equation}
2 \alpha_1 - 4 \sqrt{\frac{3}{2}} \alpha_3 + 1 = 0
\end{equation}
So we have one paramater family of Lagrangians. Now, if we calculate
the commutator of two $\eta$ transformations on the vector field
$A_\mu$, we obtain:
\begin{equation}
[ \delta_1, \delta_2] A_\mu = - 4 i e^2 (\alpha_1{}^2 + 2
\alpha_3{}^2) (\bar{\eta_2} \gamma^\nu \eta_1) A_{\nu\mu}
\end{equation}
Due to relation on the parameters $\alpha_1$ and $\alpha_3$ it is
impossible to set both parameters simultaneously equal to zero. This
means that commutator of two $\eta$ transformations always give a
translation and to proceed further on we must introduce graviton as
well. So any model with minimal gauge interaction for massive spin
3/2 particle in flat space must be a part of some (spontaneously
broken) supergravity theory. In this case all coupling constants are
related with gravitational one $k \sim \frac{1}{m_{pl}}$ and we
obtain usual for supergravity models relation betwen spin 3/2 mass,
Plank mass and gauge coupling constant $ m \sim e m_{pl}$!
Note in pass that one more example of constructive approach, namely
electromagnetic interactions for massive spin 2 particles, could be
found in \cite{KZ97}.

\section{Free massive spin 2}

Our main starting point is the gauge invariant description of free
massive spin 2 particle. As we will see part of the results obtained
crucially depend on the space-time dimension $d$ and/or the presence
or absence of cosmological term. So in what follows we will work in
space-time with arbitrary $d \ge 3$ dimension and with non-zero
(positive or negative) cosmological term. We denote metric
of such space as $g_{\mu\nu}$ and appropriate covariant derivatives
as $D_\mu$. Till the last section, where $g_{\mu\nu}$ will become
dynamical, it is just fixed background metric. Our convention for the
commutator of two covariant derivatives is:
\begin{equation}
[ D_\mu, D_\nu ] v_\alpha = \kappa ( g_{\mu\alpha} \delta_\nu{}^\beta
- \delta_\mu{}^\beta g_{\nu\alpha} ) v_\beta
\end{equation}
where $\kappa = - \frac{2 \Lambda}{(d-1)(d-2)}$, $\Lambda$ ---
cosmological term.

One of the nice features of gauge invariant description is that it
allows one to consider general case of non-zero mass and cosmological
term including all possible massless and partially massless limits
\cite{Zin01}. To describe the massive spin 2 particle in
$d$-dimensional constant curvature space we will use the following
Lagrangian:
\begin{eqnarray}
{\cal L}_0 &=& \frac{1}{2} D^\alpha h^{\mu\nu} D_\alpha h_{\mu\nu}
- \frac{1}{2} D^\alpha h^{\mu\nu} D_\mu h_{\nu\alpha} - \frac{1}{2}
(D h)^\mu (D h)_\mu + (D h)^\mu D_\mu h - \frac{1}{2} D^\mu h D_\mu h
- \nonumber \\
 && - \frac{1}{2} (D_\mu A_\nu - D_\nu A_\mu)^2 + \frac{2(d-1)}{d-2}
(D_\mu \varphi)^2 + 2m (h^{\mu\nu} D_\mu A_\nu - h (D A)) - \nonumber
\\
 && - \frac{4(d-1)c_0}{d-2} A^\mu D_\mu \varphi - 
 \frac{m^2 + \kappa(d-2)}{2} ( h^{\mu\nu} h_{\mu\nu} - h^2 ) -
\nonumber \\
 && - \frac{2(d-1)mc_0}{d-2} h \varphi + \frac{2d(d-1)m^2}{(d-2)^2}
\varphi^2 + 2\kappa(d-1) A_\mu{}^2
\end{eqnarray}
where $c_0 = \sqrt{m^2 + \kappa(d-2)}$. Here symmetric tensor
$h_{\mu\nu}$ is the main gauge field, while vector $A_\mu$ and scalar
$\varphi$ are Goldstone fields necessary for gauge invariance. To
simplify the calculations we have chosen non-canonical normalization
for kinetic terms of these fields. Also note that there is an
ambiguity in the structure of kinetic terms for the $h_{\mu\nu}$
field. Indeed, in flat space (where derivatives commute) the second
and third terms in the first line differ by total divergence only.
But in spaces with non-zero cosmological term they lead to slightly
different structure of mass-like terms and the choice we made gives
the most simple and convenient structure of these terms. It is not
hard to check that this Lagrangian is invariant under the following
local gauge transformations:
\begin{eqnarray}
\delta_0 h_{\mu\nu} &=& D_\mu \xi_\nu + D_\nu \xi_\mu +
\frac{2m}{d-2} g_{\mu\nu} \lambda \nonumber \\
\delta_0 A_\mu &=& D_\mu \lambda + m \xi_\mu \qquad 
\delta_0 \varphi = c_0 \lambda
\end{eqnarray}

Let us note, that these and all subsequent results depend crucially
on use of canonical description of massless spin 2 particles. As is
known \cite{ABGV06}, there exists alternative possibility based on
traceless symmetric tensor (or on additional Weyl symmetry).
Canonical analysis shows that right number of physical degrees of
freedom in this case is achieved due to combination of first and
second class constraints. So the relation between alternative and
canonical formulations is similar to that between, say,
non gauge invariant and gauge invariant formulations of massive spin
1 particle. Indeed, additional scalar degree of freedom (trace $h$ in
this case) promotes second class constraints to the first class ones.
As we have already note in the Introduction, it is hard to formulate
constructive approach for theories with second class constraints.
Moreover, as it was shown in \cite{ABGV06}, any attempt to give mass
to graviton in such alternative formulation unavoidably leads to the
appearance of ghosts.

Note also, that there exists a disagreement on the definition of mass
in (A)dS space. Indeed, even for the massless particles gauge
invariance requires some ``mass-like'' terms in the Lagrangians to be
present and one often tries to interpret these terms as real mass.
Gauge invariant description of massive particles gives, above all,
simple and unambiguous definition of massless limit, namely it is a
limit where Goldstone fields decouple from the main gauge field.

Let us stress the important difference between massless and massive
cases for spin 2 particle. In the massless case we have one gauge
symmetry with vector parameter $\xi_\mu$ only, while for non-zero
mass we have two gauge symmetries with vector $\xi_\mu$ and scalar
$\lambda$ parameters and only these two symmetries together could
guarantee the right number of physical degrees of freedom even after
switching on an interaction. Thus, models based on the $\xi_\mu$
invariance only like those based on $OSP(4)$ symmetry \cite{Cha03} or
models exploring $\lambda$ transformations only \cite{DW06} always
require additional checks for consistency.
One more example is a BRST construction of \cite{GS04}, where for
massive spin 2 particles authors promote gauge transformations with
vector parameter only (and introduce vector Goldstone field $u_\mu$
only). As a result, such theory describes additional scalar degrees
of freedom and all previous results point that it must be ghost. At
the same time, BRST construction for pure massive spin 2 particle
does exist \cite{Ham05}, but requires additional scalar gauge
transformations and scalar Goldstone field.

Note also that if one introduces gauge invariant under the $\lambda$
transformations derivatives:
\begin{equation}
\nabla_\mu h_{\alpha\beta} = D_\mu h_{\alpha\beta} - \frac{2m}{d-2}
A_\mu \eta_{\alpha\beta}, \qquad \nabla_\mu \varphi = D_\mu \varphi -
c_0 A_\mu
\end{equation}
then the Lagrangian could be rewritten in more simple form:
\begin{eqnarray}
{\cal L}_0 &=& \frac{1}{2} \nabla^\alpha h^{\mu\nu} \nabla_\alpha
h_{\mu\nu} - \frac{1}{2} \nabla^\alpha h^{\mu\nu} \nabla_\mu
h_{\nu\alpha} - \frac{1}{2} (\nabla h)^\mu (\nabla h)_\mu + 
(\nabla h)^\mu \nabla_\mu h - \nonumber \\
 && - \frac{1}{2} \nabla^\mu h \nabla_\mu h 
 - \frac{1}{2} (D_\mu A_\nu - D_\nu A_\mu)^2 + \frac{2(d-1)}{d-2}
(\nabla_\mu \varphi)^2 - \nonumber \\
 && - \frac{m^2 + \kappa(d-2)}{2} ( h^{\mu\nu} h_{\mu\nu} - h^2 ) -
\frac{2(d-1)mc_0}{d-2} h \varphi + \frac{2d(d-1)m^2}{(d-2)^2}
\varphi^2
\end{eqnarray}

As could be easily seen from the formulas given above in the massless
limit $m \rightarrow 0$ the Lagrangian breaks into two independent
parts. One of them (for symmetric tensor $h_{\mu\nu}$) gives usual
description of massless spin 2 particle in $(A)dS_d$ background,
while the other one (with vector $A_\mu$ and scalar $\varphi$ fields)
gives gauge invariant description of massive spin 1 particle (or sum
of massless spin 1 and spin 0 particles in a flat case). But in de
Sitter space $\kappa < 0$ there exist one more special limit $c_0
\rightarrow 0$. In this, the scalar field $\varphi$ completely
decouples, while pair $h_{\mu\nu}$, $A_\mu$ corresponds to the so
called partially massless spin 2 particle. Note that one can use
$\xi_\mu$ gauge transformation in order to set vector $A_\mu = 0$.
In this, the resulting simple Lagrangian for $h_{\mu\nu}$ will still
be invariant under the $\lambda$ transformations, provided we
supplement it with restoring gauge $A_\mu=0$ transformation with
$\xi_\mu = - \frac{1}{m} D_\mu \lambda$: 
\begin{equation}
\delta h_{\mu\nu} = - \frac{1}{m} (D_\mu D_\nu + D_\nu D_\mu)
\lambda + \frac{2m}{d-2} g_{\mu\nu} \lambda
\end{equation}
We have explicitly checked that such gauge fixed Lagrangian is
indeed invariant under this transformation.

\section{Self-interaction}

In this section we consider self-interaction of such massive spin 2
particles in linear approximation. As we have already mentioned in
the introduction, we will not make any suggestion on the structure of
gauge algebra which could stand behind this theory nor we will not
insist on any geometrical interpretation. Instead, we will
construct the most general Lorentz invariant cubic terms for the
Lagrangian as well as the most general ansatz for gauge
transformations and require that the Lagrangian will be gauge
invariant. As a result this section will be the most technical part
of the paper, but it is important to make sure that our results are
completely general.

Even for the massless particles in $(A)dS_d$ gauge invariance require
introduction of mass-like terms into the Lagrangian and appropriate
corrections to gauge transformation laws so that the structure of the
theory resembles that of massive theory in flat space. Thus, working
with general case of massive particles in $(A)dS_d$, it is convenient
to organize the calculations just by the number of derivatives. So we
start here with the Lagrangian terms with two derivatives and gauge
transformations with one derivative and analyze all possible vertexes
compatible with gauge invariance up to the lower derivatives terms.

{\bf $hhh$-vertex}. In this case the most general ansatz for gauge
transformations linear in $h$-field and containing one derivative
looks as follows:
\begin{eqnarray*}
\delta h_{\mu\nu} &=& c_1 \xi^\alpha D_\alpha h_{\mu\nu} + c_2
((Dh)_\mu \xi_\nu + (Dh)_\nu \xi_\mu) + c_3 g_{\mu\nu} (Dh)_\alpha
\xi^\alpha + \\
 &+& c_4 g_{\mu\nu} D_\alpha h \xi^\alpha + c_5 (D_\mu h_{\nu\alpha} +
D_\nu h_{\mu\alpha}) \xi^\alpha + c_6 (D_\mu h \xi_\nu + D_\nu h
\xi_\mu) + \\
 &+& c_7 (h_{\alpha\mu} D_\nu \xi^\alpha + h_{\nu\alpha} D_\mu
\xi^\alpha) + c_8 h (D_\mu \xi_\nu + D_\nu \xi_\mu) + c_9 h_{\mu\nu}
(D\xi) + \\
 &+& c_{10} (h_\mu{}^\alpha D_\alpha \xi_\nu + h_\nu{}^\alpha
D_\alpha \xi_\mu) + c_{11} g_{\mu\nu} h^{\alpha\beta} D_\alpha
\xi_\beta + c_{12} g_{\mu\nu} h (D\xi) 
\end{eqnarray*}
But due to gauge invariance of free Lagrangian the structure of this
transformations could be defined only up to arbitrary field dependent
gauge transformations (or in other words up to possible redefinitions
of $\xi_\mu$ parameter) which in this case have the form:
$$
\delta h_{\mu\nu} = d_1 [ D_\mu (h_{\nu\alpha} \xi^\alpha) + D_\nu
(h_{\mu\alpha} \xi^\alpha)] + d_2 [ D_\mu (h\xi_\nu) + D_\nu
(h\xi_\mu)]
$$
Moreover, as in all theories where interacting Lagrangian contains
the same number of derivatives as the free one, there always exists a
family of physically equivalent Lagrangians related by trivial field
redefinitions. In the case at hands, we have a possibility to make
such redefinition with four arbitrary parameters:
$$
h_{\mu\nu} \Rightarrow h_{\mu\nu} + s_1 h_\mu{}^\alpha h_{\alpha\nu}
+ s_2 h h_{\mu\nu} + s_3 g_{\mu\nu} h_{\alpha\beta}{}^2 + s_4
g_{\mu\nu} h^2
$$
We use all these freedom to bring the gauge transformations to the
following more simple form:
\begin{eqnarray*}
\delta h_{\mu\nu} &=& c_1 \xi^\alpha D_\alpha h_{\mu\nu} + c_2
((Dh)_\mu \xi_\nu + (Dh)_\nu \xi_\mu) + c_3 g_{\mu\nu} (Dh)_\alpha
\xi^\alpha + \\
 &+& c_4 g_{\mu\nu} D_\alpha h \xi^\alpha + 
 c_7 (h_{\alpha\mu} D_\nu \xi^\alpha + h_{\nu\alpha} D_\mu
\xi^\alpha) + c_9 h_{\mu\nu} (D\xi)
\end{eqnarray*}
At last, we require that the algebra of these transformations be
closed, i.e.
$$
[ \delta(\xi_1), \delta(\xi_2)] = \delta(\xi_3)
$$
Simple calculations immediately give $c_2 = c_3 = c_4 = c_9 = 0$ and
$c_7 = c_1$. In the massless case the only remaining parameter
$c_1$ corresponds to gravitational coupling constant $k \sim
1/m_{pl}$. In what follows, we set $c_1 = 2$. Thus, we obtain a very
simple final form for these gauge transformations:
\begin{equation}
\delta h_{\mu\nu} = 2 [ \xi^\alpha D_\alpha h_{\mu\nu} + 
 D_\mu \xi^\alpha  h_{\alpha\nu} +  D_\nu \xi^\alpha h_{\alpha\mu}]
\end{equation}
In this,
$$
\xi_{3\mu} = \xi_2{}^\alpha D_\alpha \xi_{1\mu} - \xi_1{}^\alpha
D_\alpha \xi_{2\mu}
$$
Certainly, these transformations look exactly like general coordinate
transformations for covariant second rank symmetric tensor, but let
us stress that in our case these are just gauge transformations for
spin 2 field living in fixed $(A)dS_d$ background. Recall also, that
concrete form of these transformations (i.e. that of covariant tensor
and not that of contravariant one or tensor density) is just a matter
of our choice. This and all our results that follows are always
defined up to possible field redefinitions of the type shown above.

Now we construct the most general Lorentz invariant cubic terms with
two derivatives:
\begin{eqnarray}
{\cal L}_{hhh} &=& h^{\mu\nu} [ a_1 D_\mu h_{\nu\alpha} (Dh)^\alpha +
 a_2 D_\mu h_{\nu\alpha} D^\alpha h + 
 a_3 D_\mu h_{\alpha\beta} D_\nu h^{\alpha\beta} +
 a_4 D_\mu h D_\nu h + \nonumber \\
 &+& a_5 D_\mu h^{\alpha\beta} D_\nu h_{\alpha\beta} +
 a_6 D_\mu h (Dh)_\nu + 
 a_7 D_\alpha h_{\mu\nu} (Dh)^\alpha +
 a_8 D_\alpha h_{\mu\nu} D_\alpha h + \nonumber \\
 &+& a_9 D_\alpha h_{\beta\mu} D^\beta h^{\alpha\nu} +
 a_{10} D_\alpha h_{\beta\mu} D^\alpha h^{\beta\nu} +
 a_{11}  (Dh)_\mu (Dh)_\nu ] + 
 h [ a_{12} D_\mu h_{\alpha\beta} D^\mu h^{\alpha\beta} + \nonumber \\
 &+& a_{13} D_\mu h D^\mu h +
 a_{14} (Dh)_\mu (Dh)^\mu +
 a_{15} (Dh)_\mu D^\mu h +
 a_{16} D_\mu h_{\alpha\beta} D^\alpha h^{\beta\mu} ]
\end{eqnarray}
As in the free case, we face an ambiguity because in a flat space not
all these terms are independent, namely up to total divergence we have
(schematically):
\begin{eqnarray*}
(a_1) &=& (a_5) + (a_9) - (a_{11}) \\
(a_2) &=& (a_6) + (a_{14}) - (a_{16})
\end{eqnarray*}
so there exists a family of equivalent Lagrangians with two arbitrary
parameters. In a constant curvature space different choices lead to
slightly different structure of mass-like $h^3$ terms and again we
use this freedom to bring these terms to the most simple form. Then
the requirement that the Lagrangian will be gauge invariant (up to
lower derivative terms given below) gives:
$$
a_1 = a_{11} = 3/2, \quad a_2 = - 3/2, \quad a_3 = - 1, \quad
a_4 = a_{15} = 1, \quad a_5 = 5/2, \quad a_6 = - 5/2, \quad
$$
$$
a_7 = a_{10} = - 2, \quad a_8 = 2, \quad a_9 = a_{12} = 1/2, \quad
a_{13} = a_{14} = a_{16} = - 1/2
$$
In this, in the non-flat space background we still have
non-compensated variations to be taken into account in the lower
derivatives orders:
\begin{eqnarray}
\delta_0 {\cal L}_{hhh} + \delta_1 {\cal L}_0 &=&
\kappa [ (3d-8) h^{\alpha\beta} D_\alpha h_{\beta\mu} -
 2(2d-5) h^{\alpha\beta} D_\mu h_{\alpha\beta} +
 2(d-3) h D_\mu h - \nonumber \\
 &-& 2(d-3) h_\mu{}^\alpha D_\alpha h +
 (3d-8) h_\mu{}^\alpha (Dh)_\alpha - 2(d-3) h (Dh)_\mu ] \xi^\mu
\end{eqnarray}

Till now all calculations are the same for massive as well as
massless spin 2 particles. Let us make here a comment on the so
called gravity reconstruction, i.e. on possibility to reconstruct
full gravity theory starting from free massless spin 2 particle in
flat or constant curvature space background \cite{Pad04}. As usual,
in all calculations of the type given above, all the variations which
are total divergence are dropped out and of course none of such terms
can be "reconstructed" in such iterations. But nothing prevent of to
add to the Lagrangian any such terms to bring the result in a
convenient form. As for the ambiguity which arise in this iterative
process and which is often related with ambiguity in the definition
of energy-momentum tensor for gravity or matter fields, we have seen
that this ambiguity clearly related with trivial field redefinitions
and up to this freedom the results are essentially unique.

{\bf $hAA$-vertex}. Here we start with the most general ansatz for
$\xi_\mu$ transformations:
\begin{eqnarray*}
\delta A_\mu &=& c_1 D_\mu A_\nu \xi^\nu + c_2 D_\nu A_\mu \xi^\nu +
c_3 (D A) \xi_\mu + \\
 && + c_4 A_\nu D_\mu \xi^\nu + c_5 A^\nu D_\nu \xi_\mu + c_6 A_\mu
(D \xi)
\end{eqnarray*}
As in the previous case there exists a freedom connected with the
possible redefinitions of field $A_\mu$ and parameter $\xi_\mu$:
$$
\delta A_\mu = d_1 D_\mu (A \xi) \qquad A_\mu \rightarrow A_\mu + s_1
h_{\mu\nu} A^\nu + s_2 h A_\mu
$$
so without lost of generality we can restrict ourselves to
$$
\delta A_\mu = c_1 D_\mu A_\nu \xi^\nu + c_2 D_\nu A_\mu \xi^\nu +
c_3 (D A) \xi_\mu
$$
Usually, then one considers an interaction of gravity with vector
fields (abelian or non-abelian) one assumes that gravitational field
$h_{\mu\nu}$ is inert under the gauge transformations of these
fields. But now vector field $A_\mu$ is just a component of massive
spin 2 field, so we have to consider all the possible gauge
transformations with $\lambda$ parameter also. Most general form
looks like:
\begin{eqnarray*}
\delta h_{\mu\nu} &=& c_4 D_{(\mu} A_{\nu)} \lambda + c_5 g_{\mu\nu}
(DA) \lambda + c_6 A_{(\mu} D_{\nu)} \lambda + c_7 g_{\mu\nu}
A^\alpha D_\alpha \lambda \\
\delta A_\mu &=& c_8 D_\mu h \lambda + c_9 (Dh)_\mu \lambda + c_{10}
h D_\mu \lambda + c_{11} h_{\mu\nu} D^\nu \lambda
\end{eqnarray*}
Once again using the freedom that exists here:
$$
\delta h_{\mu\nu} = d_2 D_{(\mu} (A_{\nu)} \lambda) \qquad
\delta A_\mu = d_3 D_\mu (h \lambda) \qquad
h_{\mu\nu} \rightarrow h_{\mu\nu} + s_3 A_\mu A_\nu + s_4 g_{\mu\nu}
A_\alpha{}^2
$$
we can reduce these transformations to the form:
\begin{eqnarray*}
\delta h_{\mu\nu} &=& c_5 g_{\mu\nu} (DA) \lambda \\
\delta A_\mu &=& c_9 (Dh)_\mu \lambda + c_{10} h D_\mu \lambda +
c_{11} h_{\mu\nu} D^\nu \lambda
\end{eqnarray*}
Now we proceed by constructing the most general Lorentz invariant
cubic terms with two derivatives:
\begin{eqnarray*}
{\cal L}_{hAA} &=& h^{\mu\nu} [ a_1 D_\mu A_\nu (DA) + a_2 D_\mu
A^\alpha D_\nu A_\alpha + a_3 D_\mu A^\alpha D_\alpha A_\nu + a_4
D^\alpha A_\mu D_\alpha A_\nu ] + \\
 && + h [ a_5 D^\alpha A^\beta D_\alpha A_\beta + a_6 D^\alpha
A_\beta D_\beta A_\alpha + a_7 (DA) (DA) ] + \\
 && + a_8 D^\alpha h^{\mu\nu} D_\alpha A_\mu A_\nu + a_9 (Dh)^\mu
D_\mu A_\alpha A^\alpha + a_{10} (Dh)^\mu D_\alpha A_\mu A^\alpha + \\
 && + a_{11} (Dh)^\mu (DA) A_\mu + a_{12} D^\mu h D_\mu A_\alpha
A^\alpha + a_{13} D^\mu h (DA) A_\mu 
\end{eqnarray*}
Not all these terms are independent here. Indeed, it is easy to check
that up to terms without derivatives:
\begin{eqnarray*}
D^\alpha h^{\mu\nu} D_\mu A_\nu A_\alpha &=& h^{\mu\nu} D_\alpha A_\nu
D_\mu A^\alpha - h^{\mu\nu} D_\mu A_\nu (DA) + (Dh)^\mu D_\alpha A_\mu
A^\alpha \\
D^\mu h D_\nu A_\mu A^\nu &=& h (DA) (DA) - h D^\alpha A^\beta
D_\alpha A_\beta + D^\mu h (DA) A_\mu \\
D^\alpha h^{\mu\nu} D_\mu A_\alpha A_\nu &=& (Dh)^\mu (DA) A_\mu +
(Dh)^\mu D_\alpha A_\mu A^\alpha - D^\alpha h^{\mu\nu} D_\mu A_\nu
A_\alpha
\end{eqnarray*}
Using this freedom and requiring that the Lagrangian will be gauge
invariant we finally get:
\begin{equation}
{\cal L}_{hAA} = c_2 h^{\mu\nu} A_{\mu\alpha} A_\nu{}^\alpha -
 \frac{c_2}{4} h A_{\mu\nu}{}^2
\end{equation}
while the only non-trivial transformation is:
\begin{equation}
\delta A_\mu = c_2 \xi^\nu A_{\nu\mu}
\end{equation}
The result obtained is of course familiar and could seems trivial,
but for what follows it is very important that linear approximation
does not fix the value of $c_2$ coupling constant. In the ``normal''
massless gravity one expects $c_2 = 2$ but as we will see later on
massive theory is possible provided $c_2 = 1$ so that Goldstone field
$A_\mu$ must have non-canonical $\xi_\mu$ transformations.

{\bf $hh\varphi$-vertex}. Now the most general ansatz for $\xi_\mu$
transformations has the form:
\begin{eqnarray*}
\delta h_{\mu\nu} &=& c_1 (D_\mu \varphi \xi_\nu + D_\nu \varphi
\xi_\mu) + c_2 g_{\mu\nu} (D_\alpha \varphi \xi^\alpha) + c_3 \varphi
( D_\mu \xi_\nu + D_\nu \xi_\mu) + c_4 \varphi g_{\mu\nu} (D\xi) \\
\delta \varphi &=& c_5 (Dh)_\mu \xi^\mu + c_6 D_\mu h \xi^\mu + c_7
h^{\mu\nu} D_\mu \xi_\nu + c_8 h (D\xi)
\end{eqnarray*}
and using one more time the freedom to make redefinitions:
$$
\delta h_{\mu\nu} = d_1 [ D_\mu(\varphi \xi_\nu) + D_\nu (\varphi
\xi_\mu)], \quad h_{\mu\nu} \rightarrow h_{\mu\nu} + s_1 \varphi
h_{\mu\nu} + s_2 \varphi g_{\mu\nu} h, \quad \varphi \rightarrow
\varphi + s_3 h^{\mu\nu} h_{\mu\nu} + s_4 h^2
$$
one can leave $c_2$, $c_5$ and $c_6$ as the only non-zero parameters.
Then writing most general cubic terms:
\begin{eqnarray*}
{\cal L}_{hh\varphi} &=& \varphi [ a_1 D^\mu h^{\alpha\beta} D_\mu
h_{\alpha\beta} + a_2 D^\mu h^{\alpha\beta} D_\alpha h_{\beta\mu} +
a_3 (Dh)^\mu (Dh)_\mu + a_4 (Dh)^\mu D_\mu h + \\
 && + a_5 D^\mu h D_\mu h ] + D^\mu \varphi [ a_6 D_\mu
h_{\alpha\beta} h^{\alpha\beta} + a_7 D_\alpha h_{\beta\mu}
h^{\alpha\beta} + a_8 (Dh)^\nu h_{\mu\nu} + \\
 && + a_9 D^\nu h h_{\mu\nu} + a_{10} D_\mu h h + a_{11} (Dh)_\mu h ]
\end{eqnarray*}
and using the fact that up to lower derivative terms we have:
$$
(a_2) = (a_3) - (a_7) + (a_8)
$$
one can check that there is no non-trivial solution for such vertex.

{\bf $AA\varphi$-vertex}. In this case one has to consider $\lambda$
transformations only, so we write:
$$
\delta A_\mu = c_1 \varphi D_\mu \lambda + c_2 D_\mu \varphi \lambda
\qquad \delta \varphi = c_3 (DA) \lambda + c_4 A^\mu D_\mu \lambda
$$
Possible field and parameter $\lambda$ redefinitions
$$
\delta A_\mu = d_1 D_\mu (\varphi \lambda), \qquad A_\mu \rightarrow
A_\mu + s_1 \varphi A_\mu, \qquad \varphi \rightarrow \varphi + s_2
A_\mu{}^2
$$
allows one to leave $c_3$ as the only non-zero parameter. Then
considering the most general cubic terms:
\begin{eqnarray*}
{\cal L}_{\varphi AA} &=& \varphi [ a_1 D^\mu A^\nu D_\mu A_\nu + a_2
D^\mu A^\nu D_\nu A_\mu + a_3 (DA) (DA) ] + \\
 && + D^\mu \varphi [ a_4 D_\mu A_\nu A^\nu + a_5 (DA) A_\mu ]
\end{eqnarray*}
and taking into account that
$$
\varphi D^\mu A^\nu D_\nu A_\mu = \varphi (DA) (DA) + D^\mu \varphi
(DA) A_\mu - D^\mu D_\nu A_\mu A^\nu
$$
we obtain the following simple Lagrangian
\begin{equation}
{\cal L}_{\varphi A A} = a_0 \varphi A_{\mu\nu}{}^2
\end{equation}
which is trivially gauge invariant.

{\bf $h\varphi\varphi$-vertex}. Here the only possible terms for
$\xi_\mu$ transformations are:
$$
\delta \varphi = c_3 D_\mu \varphi \xi^\mu + c_4 \varphi (D\xi)
$$
Moreover, using field redefinition
$$
\varphi \rightarrow \varphi + s_1 h \varphi
$$
we can leave $c_3$ as the only non-zero parameter. Then the
requirement that the Lagrangian
$$
{\cal L}_{h\varphi\varphi} = a_1 h^{\mu\nu} D_\mu \varphi D_\nu 
\varphi+ a_2 h D^\mu \varphi D_\mu \varphi + a_3 (Dh)^\mu D_\mu
\varphi \varphi + a_4 D^\mu h D_\mu \varphi \varphi
$$
will be gauge invariant gives:
$$
a_1 = - \frac{2 c_3(d-1)}{d-2}. \qquad a_2 = \frac{c_3(d-1)}{d-2},
\qquad a_3 = - a_4
$$
In this, the arbitrary parameter $a_3$ is related with one more field
redefinition (which does not change the structure of gauge
transformations)
$$
h_{\mu\nu} \rightarrow h_{\mu\nu} + s_2 g_{\mu\nu} \varphi^2
$$
So without lost of generality we can set $a_3 = 0$ and obtain:
\begin{eqnarray}
{\cal L}_{h\varphi\varphi} &=&  \frac{2(d-1)}{d-2} [ - c_3 h^{\mu\nu}
D_\mu \varphi D_\nu \varphi + \frac{c_3}{2} h (D \varphi)^2 ]
\nonumber \\
\delta \varphi &=& c_3 \xi^\mu D_\mu \varphi 
\end{eqnarray}
Again this result could seems trivial but it is important that
coupling
constant $c_3$ is not fixed yet.

{\bf $\varphi\varphi\varphi$-vertex}. In this last and simplest case
we have only one possible term in the Lagrangian, no non-trivial gauge
transformations and one possible field redefinition:
$$
{\cal L}_{\varphi^3} = a_1 \varphi D^\mu \varphi D_\mu \varphi,
\qquad \varphi \rightarrow \varphi + s_1 \varphi^2
$$
showing that this vertex is trivial one.

Let us collect together all the pieces obtained so far.
Total Lagrangian with two derivatives:
\begin{eqnarray}
{\cal L}_0 &=& {\cal L}_{hhh} + c_2 h^{\mu\nu} A_{\mu\alpha}
A_\nu{}^\alpha - \frac{c_2}{4} h A_{\mu\nu}{}^2 + a_0 \varphi
A_{\mu\nu}{}^2 + \nonumber \\
 && + \frac{2(d-1)}{d-2} [ - c_3 h^{\mu\nu} D_\mu \varphi D_\nu
\varphi + \frac{c_3}{2} h (D \varphi)^2  ]
\end{eqnarray}
while the only non-trivial gauge transformations look like:
\begin{eqnarray}
\delta h_{\mu\nu} &=& 2 [ \xi^\alpha D_\alpha h_{\mu\nu} + 
 D_\mu \xi^\alpha  h_{\alpha\nu} +  D_\nu \xi^\alpha h_{\alpha\mu}]
\nonumber \\
\delta A_\mu &=& c_2 \xi^\nu A_{\nu\mu} \qquad
\delta \varphi = c_3 \xi^\mu D_\mu \varphi 
\end{eqnarray}
Let us stress once again that parameters $c_2$ and $c_3$ are not
fixed yet.

Now we proceed to the next order. Namely, we construct the most
general cubic terms with one derivative for the Lagrangian as well as
the most general linear terms without derivatives for gauge
transformations laws. At this order there are no ambiguities related
with field redefinitions so all calculations are completely
straightforward. The resulting part of the Lagrangian has the form:
\begin{eqnarray}
{\cal L}_1 &=& m [ 4 h^{\mu\nu} (D h)_\mu A_\nu - h (D h)_\mu
A^\mu + 2 h^{\mu\nu} D_\mu h_{\nu\alpha} A^\alpha - \nonumber \\
 && - 3 h^{\mu\nu} D_\mu h A_\nu - 3 h^{\mu\nu} D_\alpha h_{\mu\nu}
A^\alpha + h D_\mu h A^\mu ] + \nonumber \\
 && + 2 c_0 b_0 h^{\mu\nu} A_\mu D_\nu \varphi - c_0 b_0 h 
(A D\varphi) + b_0 m A^\mu D_\mu \varphi \varphi
\end{eqnarray}
while additional terms to the gauge transformations are:
\begin{eqnarray}
\delta h_{\mu\nu} &=& m [ A_\mu \xi_\nu + A_\nu \xi_\mu -
\frac{4}{d-2} g_{\mu\nu} (A \xi) + \frac{6-d}{d-2} h_{\mu\nu} \lambda
] \nonumber \\
\delta A_\mu &=& m h_{\mu\nu} \xi^\nu + 2 a_0 m \varphi \xi_\mu \qquad
\delta \varphi = - c_0 c_3 (A \xi) - \frac{m c_3}{2} \varphi \lambda
\end{eqnarray}
Here $b_0 = \frac{2 c_3 (d-1)}{d-2}$, $a_0 = \frac{m(d-4)}{2 c_0
(d-2)}$, $c_2 = 1$ and $c_3 = 1 - \frac{m^2(d-4)}{2 c_0{}^2 (d-2)}$.
As could easily be seen from these formulas, in particular from the
expressions for the $a_0$ and $c_3$ parameters, in arbitrary
space-time dimension $d$ general solution exists for $c_0 \ne 0$ only.
But in $d=4$ dimensions there exists another solution with $c_0 = 0$
then scalar field $\varphi$ completely decouples which corresponds to
the self-interaction for partially massless gravity. This fact is
clearly connected with the conformal invariance of partially massless
theories namely in $d=4$ dimensions \cite{DW04}. Now we proceed by
considering general solution with $c_0 \ne 0$ in arbitrary space-time
dimension but we will comment on partially massless gravity at the
end of this section.

The last but not least part of our calculations deals with cubic
terms without derivatives in the Lagrangian. The most general form
for these terms looks like:
\begin{eqnarray}
{\cal L}_0 &=& b_1 h^{\mu\nu} h_{\nu\alpha} h^{\alpha\mu} +
b_2 h h^{\mu\nu} h_{\mu\nu} + b_3 h^3 + b_4 \varphi h^{\mu\nu}
h_{\mu\nu} + b_5 \varphi h^2 + \nonumber \\
 && + b_6 \varphi^2 h + b_7 \varphi^3 + b_8 h^{\mu\nu} A_\mu
A_\nu + b_9 h A_\mu{}^2 + b_{10} \varphi A_\mu{}^2
\end{eqnarray}
There are no any new terms for the gauge transformations here, so the
gauge invariance must be achieved with the gauge transformations
obtained at the previous orders. Indeed, it turns out to be possible
and gives:
$$
b_1 = m^2 + \frac{(7d-16)\kappa}{6}, \quad 
b_2 = - \frac{5 m^2}{4} - \frac{(3d-7)\kappa}{2}, \quad
b_3 = \frac{m^2}{4} + \frac{(2d-5)\kappa}{6},
$$
$$
b_4 = - 2 b_5 = \frac{2 c_0 m (d-1)}{d-2} - \frac{(d-4) m^3}{2 c_0
(d-2)}, \quad
b_6 = \frac{(d^2+5d-6) m^2}{2(d-2)^2} -
\frac{(3d-2)(d-4) m^4}{4c_0{}^2(d-2)^2}
$$
$$
b_7 = 2 dc_0{}^2(d^2-11d+10) + \frac{d(d+2)(d-4)m^5}{6c_0{}^3(d-2)^3}
$$
$$
b_8 = m^2 - 2(d-1)\kappa, \quad b_9 = \frac{m^2}{2} + (d-1) \kappa,
\quad b_{10} = - \frac{2c_0(d-1)m}{d-2} + \frac{(d-4)m^3}{c_0(d-2)}
$$
The dependence of these coefficients on the space-time dimension $d$
and cosmological term $\kappa$ (recall that $c_0 = \sqrt{m^2 +
(d-2)\kappa}$) appears to be rather complicated so let us give
explicit expression for non-derivative terms in $d=4$ (any way this
dimension is special for us)
\begin{eqnarray}
{\cal L}_0 &=& (m^2 + 2\kappa) [ h^{\mu\nu} h_{\nu\alpha}
h^{\alpha\mu} - \frac{5}{4} h h^{\mu\nu} h_{\mu\nu} + \frac{1}{4} h^3
] \nonumber \\
 && + 3 c_0 m \varphi h^{\mu\nu} h_{\mu\nu} - \frac{3 c_0 m}{2}
\varphi h^2 + \frac{15 m^2}{4} \varphi^2 h - \frac{3 m^3}{c_0}
\varphi^3 + \nonumber \\
 && + (m^2 - 6\kappa) h^{\mu\nu} A_\mu A_\nu +
\frac{m^2 + 6\kappa}{2} h A_\mu{}^2 - 3 c_0 m \varphi A_\mu{}^2
\end{eqnarray}

A few comments are in order.
\begin{itemize}
\item One of the most important results of our investigation is that
possibility to switch on self-interaction in linear approximation
crucially depends on the non-canonical form of gauge transformations
for vector Goldstone field $A_\mu$ (recall, that results of linear
approximation are model independent up to restriction on the number
of derivatives). Gauge transformations we obtained:
\begin{eqnarray*}
\delta h_{\mu\nu} &=& \nabla_\mu \xi_\nu + \nabla_\nu \xi_\mu + 2
(\xi^\alpha \nabla_\alpha h_{\mu\nu} + D_\mu \xi^\alpha h_{\alpha\nu}
+ D_\nu \xi^\alpha h_{\mu\alpha}) \\
\delta A_\mu &=& m \xi_\mu + \xi^\alpha A_{\alpha\mu} + m h_{\mu\nu}
\xi^\nu
\end{eqnarray*}
resembles very much the situation with massive spin 1 particles
which we discussed in section 1:
$$
\delta A_\mu{}^a = \partial_\mu \xi^a - g \varepsilon^{abc} A_\mu{}^b
\xi^c \qquad \delta \varphi^a = m \xi^a - \frac{g}{2}
\varepsilon^{abc} \varphi^b \xi^c
$$
So by analogy with spin 1 case one can suggest that there are two
possible ways to proceed beyond linear approximation. At first, if
one evade an introduction of any additional fields in the model, then
this results in highly non-linear theory with higher and higher
derivatives. Indeed, we have explicitly checked that one can not
construct quadratic approximation without introduction of such higher
derivatives terms. On the other hand, one can try to introduce
additional fields (analog of Higgs field) in order to restrict the
number of derivatives.

\item 
Note that exactly as in the free case all the terms containing
vector fields (except kinetic ones) could be "hidden" if one introduce
$\lambda$-covariant derivatives:
\begin{eqnarray}
\nabla_\mu h_{\alpha\beta} &=& D_\mu h_{\alpha\beta} - \frac{2m}{d-2}
A_\mu g_{\alpha\beta} - \frac{6-d}{d-2} m A_\mu h_{\alpha\beta}  
\nonumber \\
\nabla_\mu \varphi &=& D_\mu \varphi - c_0 A_\mu + \frac{m c_3}{2}
A_\mu \varphi
\end{eqnarray}

\item One can always use the $\xi_\mu$ and $\lambda$ local gauge
transformations to choose the gauge where $A_\mu = 0$ and $\varphi =
0$. In this, the main difference of massive and massless theories are
the cubic completion of famous Fierz-Pauli quadratic mass terms:
$$
m^2 [ h^{\mu\nu} h_{\nu\alpha} h^{\alpha\mu} - \frac{5}{4} h
h^{\mu\nu} h_{\mu\nu} + \frac{1}{4} h^3 ]
$$
It is instructive to compare this result with that of \cite{CNPT05}
(see also \cite{DR05}) obtained in a very different context. The
investigation of \cite{CNPT05} shows that allowed cubic terms must be
combination of $h^{\mu\nu} h_{\nu\alpha} h^\alpha{}_\mu - h h^{\mu\nu}
h_{\mu\nu}$ and $h^{\mu\nu} h_{\nu\alpha} h^\alpha{}_\mu - \frac{3}{2}
h h^{\mu\nu} h_{\mu\nu} + \frac{1}{2} h^3$. It is easy to see that our
terms correspond to combination with coefficients $\frac{1}{2}$, so
at this point our results agree with that of \cite{CNPT05}.

\item In the discussions of possible mass terms for gravity it is
very often assumed that the full theory could be the sum of usual
massless gravity plus some invariant mass terms, e.g. \cite{AGS02}.
If we denote effective metric as $\hat{g}_{\mu\nu}$ and assume that
in the lowest approximation $\sqrt{-\hat{g}} \simeq \sqrt{-g} + h$
and $\hat{g}^{\mu\nu} \simeq g^{\mu\nu} - 2 h^{\mu\nu}$ then we
easily obtain:
\begin{equation}
- \frac{m^2}{2} \sqrt{-\hat{g}} \hat{g}^{\mu\alpha}
\hat{g}^{\nu\beta} (h_{\mu\nu} h_{\alpha\beta} - h_{\mu\alpha}
h_{\nu\beta}) \simeq 2 m^2 [ h^{\mu\nu} h_{\nu\alpha} h^{\alpha\mu} -
\frac{5}{4} h h^{\mu\nu} h_{\mu\nu} + \frac{1}{4} h^3 ]
\end{equation}
so the structure of cubic terms does not contradict (up to a factor
2) such assumption. Nevertheless, let us stress that in the massive
case the gauge transformations of tensor $h_{\mu\nu}$ is changed so
it is hard (if at all possible) to represent massive theory as a sum
of two parts which are separately gauge invariant.

\item In the massless limit massive spin 2 particle in $(A)dS_d$
decompose into massless spin 2 particle and massive spin 1 (or
massless spin 1 and spin 0 in flat space). In this, two gauge
transformations $\xi_\mu$ and $\lambda$ are completely independent
and $\lambda$ transformation is just usual gauge transformation of
abelian vector field. But then mass $m \ne 0$ the commutator of
$\xi_\mu$ and $\lambda$ transformations gives:
\begin{eqnarray}
 [ \delta(\lambda), \delta(\xi) ] h_{\mu\nu} &=& m [ D_\mu
(\xi_\nu \lambda) + D_\nu (\xi_\mu \lambda) ] \nonumber \\
\ [ \delta(\lambda), \delta(\xi) ] A_\mu &=& m^2 \xi_\mu
\lambda
\end{eqnarray}
so if we denote a generator of $\xi_\mu$ transformation as $P_\mu$
and that of $\lambda$ one as $D$, we get commutation relation for
Weyl group $[ P_\mu, D ] \sim m P_\mu$!

\item As we have already mentioned, apart form general solution which
exists in arbitrary dimension $d$, in $dS_4$ and only in this
dimension there exist another solution with $c_0 = \sqrt{m^2 +
2\kappa} = 0$ which corresponds to self interaction of partially
massless spin 2 particle in de Sitter background. In this, scalar
field $\varphi$ completely decouples, leaving us with $h_{\mu\nu}$ and
$A_\mu$ fields (note the absence of $h^3$ terms):
\begin{eqnarray}
{\cal L}_{int} &=& {\cal L}_{hhh} + h^{\mu\nu} A_{\mu\alpha}
A_\nu{}^\alpha - \frac{1}{4} h A_{\mu\nu}{}^2 + \nonumber \\
 && + m [ 4 h^{\mu\nu} (D h)_\mu A_\nu - h (D h)_\mu
A^\mu + 2 h^{\mu\nu} D_\mu h_{\nu\alpha} A^\alpha - \nonumber \\
 && - 3 h^{\mu\nu} D_\mu h A_\nu - 3 h^{\mu\nu} D_\alpha h_{\mu\nu}
A^\alpha + h D_\mu h A^\mu ]
\end{eqnarray}
Using local $\xi_\mu$ gauge transformation one can always choose the
gauge $A_\mu = 0$. As in the free case, the resulting Lagrangian will
still be invariant under $\lambda$ transformation provided we
supplement it with the appropriate restoring gauge $A_\mu = 0$
$\xi_\mu$ transformation. Indeed, let us look at the form of these
transformations when $A_\mu = 0$:
\begin{eqnarray}
\delta h_{\mu\nu} &=& D_\mu \xi_\nu + D_\nu \xi_\mu + m g_{\mu\nu}
\lambda + 2(\xi^\alpha D_\alpha h_{\mu\nu} + D_\mu \xi^\alpha
h_{\alpha\nu} + D_\nu \xi^\alpha h_{\alpha\mu}) + m h_{\mu\nu}
\lambda \nonumber \\
\delta A_\mu &=& D_\mu \lambda + m \xi_\mu + m h_{\mu\nu} \xi^\nu
\end{eqnarray}
Then it is easy to see that compensating gauge transformation has to
be:
$$\xi_\mu \simeq - \frac{1}{m} [ D_\mu \lambda + h_{\mu\nu} D^\nu
\lambda ]$$
We have explicitly checked that the Lagrangian for the $h_{\mu\nu}$
field is indeed invariant under such combined $\lambda$ and $\xi_\mu$
transformations. In the $A_\mu = 0$ gauge the Lagrangian looks as the
linearization of usual gravity in $dS_4$ background but this holds in
the linear approximation only. We have checked that it is impossible
to proceed with quadratic approximation without introduction of
higher derivative terms and/or some other fields.
\end{itemize}

\section{Interaction with matter}

In this section we investigate possible interactions of massive spin
2 particles with matter fields of lower spins. The strategy will be
the same as in the previous section --- we construct the most general
cubic terms with no more than two derivatives in the Lagrangians and
corresponding linear terms with no more than one derivative for gauge
transformation laws with the only requirement that the Lagrangian be
gauge invariant.

\subsection{Spin 0}

We start with the simplest case of massive spin 0 particles with the
usual free Lagrangian:
\begin{equation}
{\cal L}_0 = \frac{1}{2} (\partial_\mu \pi)^2 - \frac{m_0{}^2}{2}
\pi^2
\end{equation}
All calculations here are very simple and one can easily check that
up to possible field redefinitions an interacting Lagrangian has a
form:
\begin{eqnarray}
{\cal L}_1 &=& - \frac{c_2}{2} [ h^{\mu\nu} \partial_\mu \pi
\partial_\nu \pi - \frac{1}{2} h (\partial_\mu \pi)^2 + \frac{m}{c_0}
\varphi (\partial_\mu \pi)^2 ] - \nonumber \\
 && - \frac{c_2 m_0{}^2}{4} [ h  - 4 \frac{m}{c_0} \varphi ] \pi^2
\end{eqnarray}
while gauge transformations for $\pi$ field look like:
$$
\delta \pi = c_2 \xi^\mu \partial_\mu \pi
$$
We see that vector field $A_\mu$ does not couple at all (because
$\pi$ field is inert under $\lambda$ transformations), while
coupling of scalar field $\varphi$ clearly shows an ambiguity between
flat space and massless limits. Indeed, this coupling depends on
$m/c_0 = m/\sqrt{m^2 + (d-2)\kappa}$ and we have two different
limits. On one hand, we can take a massless limit $m \rightarrow 0$
while keeping cosmological term $\kappa$ small but non-zero. In this,
scalar field $\varphi$ completely decouples so that massless limit of
massive theory agrees with purely massless theory results. On the
other hand, if we take flat space limit $\kappa \rightarrow 0$ while
keeping non-zero mass $m$, then the coupling constant for scalar
field $\varphi$ tends to 1 and does not depend on mass any more.
As a result, in the massless limit scalar field $\varphi$ does not
decouple from matter field $\pi$.

\subsection{Spin 1}

Our next example --- interaction with massive spin 1 particles. Due to
our usage of gauge invariant description this includes massless limit
as well. Thus, we introduce two fields --- vector $B_\mu$ and
Goldstone scalar $\pi$ and start with the free Lagrangian which has
its own gauge invariance:
\begin{eqnarray}
{\cal L}_0 &=& - \frac{1}{4} B_{\mu\nu}{}^2 + \frac{1}{2} (D_\mu
\pi)^2 - m_1 B^\mu D_\mu \pi + \frac{m_1{}^2}{2} B_\mu{}^2 \nonumber
\\
\delta B_\mu &=& \partial_\mu \tilde{\lambda}, \qquad
\delta \pi = m_1 \tilde{\lambda}
\end{eqnarray}
An investigation of possible $hBB$ and $h\pi\pi$ vertexes with two
derivatives goes exactly the same way as that of $hAA$ and 
$h\varphi\varphi$ vertexes in the previous section, so we will not
repeat these calculations here and write the corresponding part of
Lagrangian:
\begin{eqnarray}
{\cal L}_1 &=&  \frac{c_1}{2} h^{\mu\nu} B_{\mu\alpha} B_{\nu\alpha}
- \frac{c_1}{8} h B_{\mu\nu}{}^2 - \frac{c_2}{2} h^{\mu\nu} D_\mu \pi
D_\nu \pi + \frac{c_2}{4} h (D_\mu \pi)^2 + \nonumber \\
 && + a_0 \varphi (D_\mu \pi)^2 + a_1 \varphi B_{\mu\nu}{}^2 + a_2 \pi
A^{\mu\nu} B_{\mu\nu}
\end{eqnarray}
which is invariant under the following transformations:
\begin{equation}
\delta B_\mu = c_1 \xi^\nu B_{\nu\mu}, \qquad \delta \pi = c_2
\xi^\mu D_\mu \pi
\end{equation}
Now we proceed by adding all possible lower derivatives terms to the
Lagrangian and corresponding terms to gauge transformations. This
procedure results in:
\begin{eqnarray}
{\cal L}_2 &=& m_1 c_1 [ h^{\mu\nu} B_\mu D_\nu \pi - \frac{1}{2} h
(B D\pi) + \frac{m}{c_0} \varphi (B D\pi) ] - \nonumber \\
 && - \frac{m_1{}^2 c_1}{2} [ h^{\mu\nu} B_\mu B_\nu - \frac{1}{2} h
B_\mu{}^2 + \frac{m}{c_0} \varphi B_\mu{}^2 ]
\end{eqnarray}
with the only new term in gauge transformations:
$$
\delta' \pi = - c_1 m_1 \xi^\mu B_\mu
$$
In this, gauge invariance requires $c_1 = c_2$, $a_0 = \frac{m}{c_0}$,
$a_1 = \frac{c_1 m (d-4)}{4 c_0 (d-2)}$, $a_2 = 0$. Once again, our
results show an ambiguity betwen flat space and massless limits. Only
for massless vector field in $d = 4$ dimensions this ambiguity is
absent.

\subsection{Spin 1/2}

Our last example --- interaction with massive spin 1/2 particles
(recall that we prefer to work with Majorana spinors). As is well
known, to describe spinor fields living in curved background one has
to use first order formalism in terms of "tetrad" $e_\mu{}^a$  and
Lorentz connection so that $\gamma_\mu = e_{\mu a} \gamma^a$ and
\begin{equation}
[ D_\mu, D_\nu ] \chi = \frac{1}{4} R_{\mu\nu}{}^{ab} \sigma_{ab}
\chi = \frac{\kappa}{2} \sigma_{\mu\nu} \chi
\end{equation}
In this, to describe an interaction of spinor fields with our massive
spin 2 particles it is also convenient to use first order formulation
of such particles \cite{Zin03a} in terms of three pairs of fields:
$(e_{\mu a},\omega_\mu{}^{ab})$, $(A_\mu, F^{ab})$ and $(\varphi,
\pi^a)$. We will not reproduce Lagrangian for such formulation here
(it could easily be found in \cite{Zin03a}) because all we need here
is the structure of gauge transformations:
\begin{eqnarray}
\delta h_{\mu a} &=& D_\mu \xi_a + \eta_{\mu a} + \frac{m}{2}
e_{\mu a} \lambda \nonumber \\
\delta \omega_\mu{}^{ab} &=& D_\mu \eta^{ab} - \frac{c_0{}^2}{2}
(e_\mu{}^a \xi^b - e_\mu{}^b \xi^a) \\
\delta A_\mu &=& D_\mu \lambda + m \xi_\mu \qquad \delta \varphi =
c_0 \lambda \nonumber
\end{eqnarray}
where apart from $\xi^a$ and $\lambda$ gauge transformations we have
now one more transformation with parameter $\eta_{ab} = - \eta_{ba}$. 
Again we start with the free Lagrangian for massive spinor $\chi$:
\begin{equation}
{\cal L}_0 = \frac{i}{2} \bar{\chi} \gamma^\mu D_\mu \chi -
\frac{m_{1/2}}{2} \bar{\chi} \chi
\end{equation}
and construct the most general cubic terms (this time with no more
than one derivative) compatible with all gauge symmetries. We obtain
for interaction Lagrangian:
\begin{eqnarray}
{\cal L}_1 &=& - \frac{i}{2} \bar{\chi} h^{\mu\nu} \gamma_\mu D_\nu
\chi + \frac{i}{2} \bar{\chi} h \gamma^\mu D_\mu \chi - \frac{i}{8}
\bar{\chi} \gamma^\mu \omega_\mu{}^{ab} \sigma_{ab} \chi - \nonumber
\\
 && - \frac{m_{1/2}}{2} (h - \frac{m}{2 c_0} \varphi) \bar{\chi}
\chi
\end{eqnarray}
while for gauge transformations we get:
\begin{equation}
\delta \chi = \xi^\mu D_\mu \chi - \frac{1}{4} (\sigma \eta) \chi -
\frac{3m}{4} \lambda \chi 
\end{equation}
In this case an ambiguity between flat space and massless limits
exists for massive $m_{1/2} \ne 0$ spinor field only, while the
results for massless spinor agree with that of purely massless
gravity.

\section{Interaction with gravity}

There exist well known difficulties one faces in any attempt to
construct interacting theory for a collection of massless spin 2
particles \cite{Wald86a,Wald86b,BDGH00,Bou02,Anc03}. For example, for
the case of just two massless spin 2 particles there are only two
possibilities. One of them corresponds to two copies of usual
gravities completely independent of each other, while the other
possibility which does have interaction requires that one of these
particles has wrong sign of kinetic terms and be a ghost. Moreover,
there are examples of theories of such kind, where one of spin 2
particles is massive, coming from higher derivative gravity models
e.g. \cite{MS02}. Also, as is well known, there are examples of
consistent theories with infinite number of massive spin 2 particles
in Kaluza-Klein models, but it could be shown \cite{Reu88,Hoh05} that
it does not contradict with general results of \cite{Wald86a,Wald86b}.

But all this still leaves a possibility to construct a consistent
theory where one massless spin 2 particle interacts with one or
several massive spin 2 ones. In this section we investigate the case
of one massless and one massive particles. In this, there are two
possible cubic vertexes (besides self-interaction). We have checked
(though we will not present details of these calculations here) that
it is impossible to construct linear approximation with two massless
and one massive particles (once again up to the restriction on the
number of derivatives). As for the other case, i.e. interaction of
massive spin 2 particles with usual gravity, it is not hard to
constract general covariant vertex with arbitrary number of massless
fields but still bi-linear in massive fields, as we are going to
demonstrate.

First of all, let us note that in this section metric $g_{\mu\nu}$ is
not just a fixed background any more, but it is a dynamical field
with its own equation of motion:
\begin{equation}
R_{\mu\nu} - \frac{1}{2} g_{\mu\nu} R + \frac{\kappa (d-1)(d-2)}{2}
g_{\mu\nu} = 0
\end{equation}
As usual in gravity, we will assume that connection is metric
compatible $D_\alpha g_{\mu\nu} = 0$ and we have usual identities:
\begin{eqnarray*}
D^\mu R_{\mu\nu,\alpha\beta} &=& D_\alpha R_{\beta\nu} - D_\beta
R_{\alpha\nu} \\
D^\mu R_{\mu\nu} &=& \frac{1}{2} D_\nu R
\end{eqnarray*}

As in the case of massive particle living in constant curvature
background, it is convenient to organize the calculations by the
number of derivatives. So we start with the sum of kinetic terms of
our three fields $h_{\mu\nu}$, $A_\mu$ and $\varphi$:
\begin{eqnarray}
{\cal L}_2 &=& \frac{1}{2} D^\alpha h^{\mu\nu} D_\alpha h_{\mu\nu} -
D^\alpha h^{\mu\nu} D_\mu h_{\nu\alpha} + (D h)^\mu D_\mu h -
\frac{1}{2} D^\mu h D_\mu h - \nonumber \\
 && - \frac{1}{2} (D_\mu A_\nu - D_\nu A_\mu)^2 +
\frac{2(d-1)}{(d-2)} D^\mu \varphi D_\mu \varphi
\end{eqnarray}
and corresponding gauge transformations with one derivative:
\begin{equation}
\delta_1 h_{\mu\nu} = D_\mu \xi_\nu + D_\nu \xi_\mu, \qquad \delta_1
A_\mu = D_\mu \lambda
\end{equation}
Covariant derivatives do not commute and, as a result, this
Lagrangian is not invariant under these gauge transformations. But
gauge invariance could be restored (up to lower derivative terms we
will take into account later) if one adds to the Lagrangian:
\begin{equation}
\Delta {\cal L} = - 2 R^{\mu\nu} h_{\mu\alpha} h_{\nu\alpha} +
R^{\mu\nu} h_{\mu\nu} h + \frac{1}{2} R h^{\alpha\beta}
h_{\alpha\beta} - \frac{1}{4} R h^2
\end{equation}
and requires that metric $g_{\mu\nu}$ has non-trivial transformation
\begin{equation}
\delta_1 g_{\alpha\beta} = 2 (D_\mu h_{\alpha\beta} - D_\alpha
h_{\beta\mu} - D_\beta h_{\alpha\mu}) \xi^\mu
\end{equation}
Now we proceed with lower derivative terms. The part of the
Lagrangian with one derivatives turns out to be:
\begin{equation}
{\cal L}_1 = 2m [ h^{\mu\nu} D_\mu A_\nu - h (D A) ] -
\frac{4(d-1)c_0}{(d-2)} A^\mu D_\mu \varphi
\end{equation}
while corrections to gauge transformation look like:
\begin{equation}
\delta_0 h_{\mu\nu} = \frac{2m}{(d-2)} g_{\mu\nu} \lambda, \qquad
\delta_0 A_\mu = m \xi_\mu, \qquad \delta_0 \varphi = c_0 \lambda
\end{equation}
Also this requires additional modification of $g_{\mu\nu}$ gauge
transformations:
\begin{equation}
\delta_0 g_{\alpha\beta} = 2m [ A_\alpha \xi_\beta + A_\beta
\xi_\alpha - \frac{2}{(d-2)} g_{\alpha\beta} (A \xi) ] -
\frac{2m(d-4)}{(d-2)} h_{\alpha\beta} \lambda
\end{equation}
At last, we must add possible terms without derivatives, the most
general form being:
\begin{equation}
{\cal L}_0 = - \frac{b_1}{2} h^{\mu\nu} h_{\mu\nu} + \frac{b_2}{2} h^2
+ b_3 h \varphi + b_4 \varphi^2 + b_5 A_\mu{}^2
\end{equation}
There are no any additional corrections to gauge transformations at
this level, so the gauge invariance must be achieved with the ones we
already have. And indeed it turns out to be possible giving:
$$
b_1 = m^2 + \kappa(d-1)(d-2), \qquad 
b_2 = m^2 + \frac{\kappa}{2}(d-1)(d-2)
$$
$$
b_3 = - \frac{2(d-1)m c_0}{(d-2)}, \qquad
b_4 = \frac{2m^2 d (d-1)}{(d-2)^2}, \qquad
b_5 = 2 \kappa (d-1)
$$
This time also there is a possibility greatly simplify all the
expressions by introduction of $\lambda$-covariant derivatives:
\begin{equation}
\nabla_\mu h_{\alpha\beta} = D_\mu h_{\alpha\beta} - \frac{2m}{(d-2)}
A_\mu g_{\alpha\beta}, \qquad \nabla_\mu \varphi = D_\mu \varphi -
c_0 A_\mu
\end{equation}
In this, total Lagrangian could be rewritten in the form:
\begin{eqnarray}
{\cal L} &=& \frac{1}{2} \nabla^\alpha h^{\mu\nu} \nabla_\alpha
h_{\mu\nu} - \nabla^\alpha h^{\mu\nu} \nabla_\mu h_{\nu\alpha} + 
(\nabla h)^\mu \nabla_\mu h - \frac{1}{2} \nabla^\mu h \nabla_\mu h -
\nonumber \\
 && - \frac{1}{2} (D_\mu A_\nu - D_\nu A_\mu)^2 +
\frac{2(d-1)}{(d-2)} \nabla^\mu \varphi \nabla_\mu \varphi - \nonumber
\\
&& - 2 R^{\mu\nu} h_{\mu\alpha} h_{\nu\alpha} +
R^{\mu\nu} h_{\mu\nu} h + \frac{1}{2} R h^{\alpha\beta}
h_{\alpha\beta} - \frac{1}{4} R h^2 - \nonumber \\
 && - \frac{b_1}{2} h^{\mu\nu} h_{\mu\nu} + \frac{b_2}{2} h^2
+ b_3 h \varphi + b_4 \varphi^2
\end{eqnarray}
while gauge transformations for the metric field $g_{\mu\nu}$ look
like:
\begin{equation}
\delta g_{\alpha\beta} = 2 (\nabla_\mu h_{\alpha\beta} - \nabla_\alpha
h_{\beta\mu} - \nabla_\beta h_{\alpha\mu}) \xi^\mu +
\frac{2m(d-4)}{(d-2)} [ g_{\alpha\beta} (A \xi) - h_{\alpha\beta}
\lambda ]
\end{equation}
Thus, in this approximation an interaction of massive spin 2
particles with gravity exists in any space-time dimension $d \ge 3$
(though from the last equation one can see that $d=4$ case is also
special here). In particular, nothing prevent us to consider the $c_0
\rightarrow 0$ limit, i.e. interaction of partially massless spin 2
particle with gravity. But as in the case of self-interaction, as we
have explicitly checked, if one tries to proceed with terms quartic
in massive fields than one will find that higher derivatives
interactions and/or some additional fields are necessary. It is
instructive to compare our results here with the investigations of
massive spin 2 particle in gravitational background
\cite{BKP99,BGKP99,BGP00}. In general, results are similar, but the
structure of $Rhh$ terms is slightly differrent.

\section*{Conclusion}

We hope that the main lesson from our paper is that constructive
approach based on gauge invariant description of massive high spin
particles does allow one efficiently investigate possible
interactions of such particles. It is important that due to
peculiarity of linear approximation the results obtained for any
particle are completely model independent and do not depend on the
presence of any other fields in the theory. In particular, an
impossibility to construct an interaction in linear approximation
means that such interaction does not exist at all. One of the
examples of such ``no-go'' results is the absence of self-interaction
for partially massless spin 2 particles in $d \ne 4$ dimensions.

One of the striking and unexpected results is that the
existence of self-interaction for massive spin 2 particles
crucially depends on non-canonical form of gauge transformations for
Goldstone $A_\mu$ field. By analogy with massive spin 1 case, one can
assume that in full interacting theory (if it exists at all) one must
deal with non-linear realization of $\xi_\mu$ symmetry with higher
and higher derivatives. We are not aware on any works on non-linear
realization of space-time symmetries where something similar appears. 

In the investigation of the massive spin 2 particle interacting with
matter (i.e. spin 0, 1, and 1/2 particles) out results clearly show
the ambiguity betwen flat space and massless limits which reveals
itself through the dependence of scalar Goldstone field $\varphi$
coupling constant on mass and cosmological constant. Let us stress
once again that this results are also model independent.

Throughout the paper we restrict ourselves with interaction terms
with no more than two derivatives in the Lagrangians (and
correspondingly no more than one derivative in gauge transformations).
But many of our results clearly show that to construct full
interacting theory beyond linear approximations one unavoidably will
have to introduce higher derivatives interactions. But such higher
derivatives interactions could, in principle, change the results
obtained here for linear approximation. This question deserves
further study.

\end{document}